\documentclass{article}[12pt]

\usepackage{amsmath}
\usepackage{amssymb}

\newtheorem{thm}{Theorem}[section]

\newtheorem{lem}[thm]{Lemma}

\newtheorem{rem}[thm]{Remark}
\numberwithin{equation}{section}

\begin{document}

\begin{center}

 Relations among various versions
\\
 of the Segal-Bargmann transform

\end{center}

\vskip .5cm

\centerline{Stephen Bruce Sontz\footnote{Research partially
supported by CONACYT (Mexico) project 49187.}}

\centerline{Centro de Investigaci\'on en Matem\'aticas, A.C. (CIMAT)}

\centerline{Guanajuato, Mexico}

\centerline{email: sontz@cimat.mx}

\vskip .8cm \noindent

\begin{abstract}
We present various relations among Versions~$A$, $B$ and $C$
of the Segal-Bargmann transform.
We get results for the Segal-Bargmann transform associated to
a Coxeter group acting on a finite dimensional Euclidean space.
Then analogous results are shown for the Segal-Bargmann transform
of a connected, compact Lie group for all except one of the identities established
in the Coxeter case.
A counterexample is given to show that the remaining identity from the Coxeter case
does not have an analogous identity for the Lie group case.
A major result is that in both contexts the Segal-Bargmann transform for Version~$C$
is determined by that for Version~$A$.
\end{abstract}



\section{A Brief Introduction}

We recall quickly some notations and definitions from \cite{SBS2}.
Many definitions and details are not presented here.
We also advise the reader that our normalizations are not standard.

A root system is a certain finite subset $\mathcal{R}$ of nonzero vectors
of $\mathbb{R}^N$ where $N \ge 1$ is an integer.
It turns out that the finite set of reflections associated to these vectors
(orthogonal reflection in the hyperplane perpendicular to each vector)
generates a finite subgroup, known as the Coxeter group, of the orthogonal
group of $\mathbb{R}^N$.
A multiplicity function is a function $\mu : \mathcal{R} \to \mathbb{C}$
invariant under the action of the Coxeter group.
We always will assume that the multiplicity function satisfies $\mu \ge 0$.
This condition is sufficient for the existence of the Segal-Bargmann
spaces considered and for the various properties that we shall use.

We will take $t > 0$ (Planck's constant) fixed throughout this paper.

We will use the holomorphic Dunkl kernel function
$E_\mu : \mathbb{C}^N \times \mathbb{C}^N \to \mathbb{C}$, which
for all $z,w \in \mathbb{C}^N$ satisfies
$E_\mu (z,w) = E_\mu (w,z)$, $E_\mu(z,0) = 1$
and $E_\mu (z^* , z) \ge 0$ among many other properties.
When $\mu \equiv 0 $, we have $E_\mu(z,w) = e^{z \cdot w}$.

We will also be using the analytic continuation of the Dunkl heat kernel,
which is given for $z,w \in \mathbb{C}^N$ by
\begin{equation}
\label{rho_def}
      \rho_{\mu, t} (z,w) = e^{-(z^2 + w^2)/2t}
           E_\mu \left(  \dfrac{z}{t^{1/2}}, \dfrac{w}{t^{1/2}} \right).
\end{equation}
This kernel arises in the solution of the initial value problem of the heat
equation associated with the Dunkl Laplacian operator.
(See \cite{RO1}.)

We next define the kernel functions of the versions of the Segal-Bargmann transform
associated to a Coxeter group
for $z \in \mathbb{C}^N$ and $q \in \mathbb{R}^N$ by
\begin{equation}
\label{define_A}
A_{\mu, t} (z,q) := e^{-z^2/2t - q^2/4t}
      E_\mu \left(  \dfrac{z}{t^{1/2}}, \dfrac{q}{t^{1/2}} \right)
\end{equation}
and
\begin{equation}
\label{define_B}
   B_{\mu, t} (z,q) := \dfrac{ \rho_{\mu, t}(z,q) }{ \rho_{\mu, t} (0,q) }
\end{equation}
and
\begin{equation}
\label{define_C}
   C_{\mu, t} (z,q) := \rho_{\mu, t}(z,q).
\end{equation}
See \cite{BA}, \cite{FO} and \cite{SE} for the origins of this theory in the
case $\mu \equiv 0$.

The versions of the Segal-Bargmann transform are given as follows.
(See \cite{SBSBO}, \cite{HA}, \cite{SISO}, \cite{SO} and \cite{SBS2}.)
Versions~$A$ and $C$ are defined by
$$
  A_{\mu, t} f(z) := \int_{\mathbb{R}^N} \mathrm{d}\omega_{\mu, t}(q) \, A_{\mu, t}(z,q) f(q)
$$
and
$$
 C_{\mu, t} f(z) := \int_{\mathbb{R}^N} \mathrm{d}\omega_{\mu, t}(q) \, C_{\mu, t}(z,q) f(q)
$$
respectively, where $z \in \mathbb{C}^N$, $f \in L^2( \mathbb{R}^N, \omega_{\mu, t} )$ and
$\omega_{\mu, t} $ is the density of a measure on $\mathbb{R}^N$.
Version~$B$ is defined by
$$
  B_{\mu, t} f(z) := \int_{\mathbb{R}^N} \mathrm{d}m_{\mu, t}(q) \, B_{\mu, t}(z,q) f(q),
$$
where $z \in \mathbb{C}^N$, $f \in L^2( \mathbb{R}^N, m_{\mu, t} )$ and
$m_{\mu, t} $ is the density of a measure on $\mathbb{R}^N$.

Associated to these versions
there are reproducing kernel Hilbert spaces of holomorphic functions
$f : \mathbb{C}^N \to \mathbb{C}$, denoted $\mathcal{A}_{\mu, t}$,
$\mathcal{B}_{\mu, t}$ and $\mathcal{C}_{\mu, t}$ respectively,
such that
\begin{gather*}
A_{\mu, t} : L^2( \mathbb{R}^N, \omega_{\mu, t}) \to \mathcal{A}_{\mu, t}
\\
B_{\mu, t} : L^2( \mathbb{R}^N, m_{\mu, t}) \to \mathcal{B}_{\mu, t}
\\
C_{\mu, t} : L^2( \mathbb{R}^N, \omega_{\mu, t}) \to \mathcal{C}_{\mu, t}
\end{gather*}
are unitary isomorphisms.
It turns out that  $\mathcal{A}_{\mu, t} =\mathcal{B}_{\mu, t}$ as Hilbert spaces.

The holomorphic function
$\rho_{\mu, t} : \mathbb{C}^N \times \mathbb{C}^N \to \mathbb{C}$ in our
opinion is not a fundamental object.
Rather we view $\sigma_{\mu, t} (q) := \rho_{\mu, t}(0,q) = e^{-q^2/2t}$
for $q \in \mathbb{R}^N$ as the fundamental Dunkl heat kernel,
even though it does not depend on $\mu$.
Then this one-variable kernel $\sigma_{\mu, t} : \mathbb{R}^N \to (0,\infty)$
gives rise to the two-variable kernel
\begin{equation}
\label{two_var_kernel}
 \rho_{\mu, t} : \mathbb{R}^N \times \mathbb{R}^N \to \mathbb{R}
\end{equation}
using a generalized (or Dunkl) translation operator, denoted $\mathcal{T}_{\mu, x}$,
via the equation
$$
   \rho_{\mu, t}( x, q ) = \mathcal{T}_{\mu, x} \sigma_{\mu, t} (q)
$$
for all $x, q \in \mathbb{R}^N$.
(See \cite{SBS2} for more details, including definitions and proofs.)
In our conventions,
note that for $\mu \equiv 0$ we have $\rho_{0, t} (x,q) = \sigma_{0,t}(q-x)$.
Finally the function
$\rho_{\mu, t} : \mathbb{C}^N \times \mathbb{C}^N \to \mathbb{C}$ is
obtained from (\ref{two_var_kernel}) by analytic continuation.

For more details about this background material
see references \cite{SBSBO}, \cite{DX}, \cite{RO1}, \cite{RO2} and \cite{SBS2},
while for other related research in Segal-Bargmann analysis
see \cite{HI}, \cite{HZ},  \cite{OZ}, \cite{OO}, \cite{SBS1} and \cite{SBS3}.

\section{Coxeter group case}

We proved the following relation between the kernel functions
for the $A$ Version and the $C$ Version of the Segal-Bargmann transform
associated to a Coxeter group in \cite{SBS2}, namely,
\begin{equation}
\label{C-A2}
C_{\mu,t}(z,q) = A_{\mu,t} (0,q) A_{\mu,t}  (z,q)
\end{equation}
for $z \in \mathbb{C}^N $ and $q \in \mathbb{R}^N$.
The reader can readily verify this using the definitions in the previous section.
As an immediate consequence we have this identity:
\begin{eqnarray}
 C_{\mu, t} \psi (z) &=& \int_{ \mathbb{R}^N } \mathrm{d} \omega_{\mu, t} (q) \,
C_{\mu,t}(z,q) \psi (q) \nonumber
\\
&=& \int_{ \mathbb{R}^N } \mathrm{d} \omega_{\mu, t} (q) \,
A_{\mu,t}(z,q) A_{\mu,t} (0,q)  \psi (q)
\label{MU}
\end{eqnarray}
for all $\psi \in L^2 ( \mathbb{R}^N, \omega_{\mu, t} )$
and all $z \in \mathbb{C}^N$.

So, we have represented the unitary operator $C_{\mu, t} $ as the composition of two operators:
the first is the operator (denoted by $M_t$) of multiplication by the bounded function
$ A_{\mu,t} (0,q) =  e^{-q^2/4t} $,
and the second is the unitary operator $A_{\mu, t} $.
In other words we can write (\ref{MU}) as
\begin{equation}
\label{Cfactorized}
C_{\mu, t} = A_{\mu, t} M_t.
\end{equation}
As far as we are aware this representation is new, even in the case when $\mu \equiv 0 $.
The boundedness of the function $A_{\mu, t}(0,q) $, where $q \in \mathbb{R}^N$, is essential
since this gives us that $M_t$ is an operator from $L^2 ( \mathbb{R}^N, \omega_{\mu, t} )$ to \textit{itself}.
Therefore, the second operator $A_{\mu, t}$ in (\ref{Cfactorized})
is acting on the space where it is a unitary operator.
Moreover, the operator norm of $M_t$ satisfies
$||M_t|| = \sup_{q \in \mathbb{R}^N} (e^{-q^2/4t}) =1$.

Though this representation of $C_{\mu, t}$ is similar to a Toeplitz operator,
it is decidedly different.
Here we have multiplication by a bounded function followed by a specific unitary operator,
while a Toeplitz operator is multiplication by a bounded function followed by a specific
projection operator.

In the case $\mu \equiv 0$, it is known that
$\mathcal{C}_{\mu, t} \subset \mathcal{A}_{\mu, t}$, a bounded inclusion.
When $\mu \equiv 0$ these two reproducing kernel Hilbert spaces can alternatively
be defined in terms of measures on $\mathbb{C}^N$.
Then the bounded inclusion follows for example from the formulas for these measures.
(See \cite{HA2}, p.~51, where the formulas given there for these measures
for $N=1$ also hold for $N > 1$.)
The generalization of this to the present context is the next result.

\begin{thm}
We have the contractive (in particular, bounded) inclusion
$$
 \mathcal{C}_{\mu, t} \subset \mathcal{A}_{\mu, t}.
$$
\end{thm}
\textbf{Proof:}
Using (\ref{Cfactorized}), we have that
$$
\mathcal{C}_{\mu, t} = \mathrm{Ran} (C_{\mu, t}) = \mathrm{Ran} (A_{\mu, t} M_t)
\subset \mathrm{Ran} (A_{\mu, t}) = \mathcal{A}_{\mu, t},
$$
which is the inclusion we wish to prove.

We next note that the inclusion map is equal to $ A_{\mu, t} M_t (C_{\mu, t})^{-1} $,
since this acts as the identity on its domain $ \mathcal{C}_{\mu, t} $ and
has codomain $\mathcal{A}_{\mu, t}$.
Therefore the inclusion map
$\iota : \mathcal{C}_{\mu, t} \hookrightarrow \mathcal{A}_{\mu, t}$,
being the composition of two bounded operators, is bounded.
Its operator norm satisfies
$$
       ||\iota || = ||A_{\mu, t} M_t (C_{\mu, t})^{-1} ||
\le ||A_{\mu, t}|| \, || M_t || \, ||(C_{\mu, t})^{-1} || =1,
$$
exactly what one requires of an inclusion for it to be contractive.
\hfill $\blacksquare$

\begin{rem}
Using the different normalizations in \cite{HA2} this inclusion is bounded,
but not contractive.
\end{rem}
\noindent
Even though equation (\ref{C-A2}) immediately implies
for $z \in \mathbb{C}^N $ and $q \in \mathbb{R}^N$ that
\begin{equation}
\label{AC}
A_{\mu,t}  (z,q) = \dfrac{C_{\mu,t}(z,q)}{A_{\mu,t} (0,q)},
\end{equation}
this factorization of $A_{\mu,t}  (z,q)$ is not very useful, since
$(A_{\mu,t} (0,q))^{-1} = e^{q^2/4t}  $ is not a bounded function of $q$.
So we are not able to prove the opposite inclusion
$\mathcal{C}_{\mu, t} \supset \mathcal{A}_{\mu, t}$
using (\ref{AC}).
Actually, we have the following.
\begin{thm}
The complementary set
 $\mathcal{A}_{\mu, t} \setminus \mathcal{C}_{\mu, t} $ is non-empty,
that is, there exists $f \in \mathcal{A}_{\mu, t}$ such that
$f \notin \mathcal{C}_{\mu, t}$.
\end{thm}
\textbf{Proof:}
It is known (see \cite{SBS3}) that $\mathcal{C}_{\mu, t}$ is a reproducing kernel Hilbert
space with reproducing kernel function
$$
          L_{\mu, t} (z,w) = c \rho_{\mu, 2t} (z^* , w)
$$
for all $z, w \in \mathbb{C}^N$, where the constant $c>0$
is not important for us now.
So any $f \in \mathcal{C}_{\mu, t}$ satisfies the usual pointwise
bound for a reproducing kernel Hilbert
space, namely
$$
      |f(z)| \le ( L_{\mu, t} (z,z) )^{1/2} \, || f ||_{\mathcal{C}_{\mu, t}}
$$
for all $z \in \mathbb{C}^N$.
Next we use the definition of the Dunkl heat kernel to calculate
$$
      L_{\mu, t} (z,z) = c \rho_{\mu, 2t} (z^* , z) = c  e^{-((z^*)^2 + z^2)/4t}
           E_\mu \left(  \dfrac{z^*}{(2t)^{1/2}}, \dfrac{z}{(2t)^{1/2}} \right).
$$
We write $z = x + iy$ with $x,y \in \mathbb{R}^N$ and so get
$(z^*)^2 + z^2 = 2 Re(z^2) = 2(x^2 - y^2)$.
We then use the estimates (see \cite{RO2})
$$
 0 \le E_\mu \left(  \dfrac{z^*}{(2t)^{1/2}}, \dfrac{z}{(2t)^{1/2}} \right) \le
          e^{ ||z||^2 / 2t } = e^{ (x^2 + y^2) / 2t}
$$
to conclude that
$$
    |f(z)| \le c^{1/2} e^{-(x^2 - y^2)/4t} e^{ (x^2 + y^2) / 4t}
\, || f ||_{\mathcal{C}_{\mu, t}} =
c^{1/2} e^{y^2 / 2t} \, || f ||_{\mathcal{C}_{\mu, t}}.
$$
In particular, it follows that $f$ restricted to ${\mathbb{R}^N}$ is a bounded function
for every $f \in \mathcal{C}_{\mu, t}$.
This implies that the only holomorphic polynomials in $\mathcal{C}_{\mu, t}$
are the constants.
But we know that $p \in \mathcal{A}_{\mu, t} $ for \textit{all} holomorphic polynomials $p$.
(See \cite{SBSBO}.)
And this shows that
 $\mathcal{A}_{\mu, t} \setminus \mathcal{C}_{\mu, t} $ is non-empty.
\hfill $\blacksquare$

However, we shall see later on in the next section
 that the relation between the Version~$A$ and the
Version~$C$ Segal-Bargmann spaces is different in the
case of compact Lie groups.
But first, we present some relations in the Coxeter context
among Versions $A$, $B$ and $C$ of the Segal-Bargmann transform
and the Dunkl heat kernel $\rho_{\mu, t}$ restricted to $\mathbb{R}^N$,
that is, $\sigma_{\mu, t} (q) = \rho_{\mu, t} (0,q)$ for  $q \in \mathbb{R}^N$.
\begin{thm}
\label{Coxeter_ids}
For $q \in \mathbb{R}^N$ and $z \in \mathbb{C}^N$ we have the identities
\begin{eqnarray}
\label{first_mu_id}
  B_{\mu, t} (z,q) &=& A_{\mu, t} (z,q) \, / \, A_{\mu, t} (0,q)
\\
\label{third_mu_id}
   \rho_{\mu, t}(z,q) = C_{\mu,t}(z,q) &=& A_{\mu,t} (0,q) A_{\mu,t}  (z,q)
\\
\label{second_mu_id}
\sigma_{\mu, t} (q) = \rho_{\mu, t} (0, q) &=& ( A_{\mu, t} (0,q) )^2
\\
\label{Euclidean_id}
 C_{\mu,t}(2z,q) &=& A_{\mu,2t} (2z,0) A_{\mu,t/2}  (z,q)
\end{eqnarray}
which tell us that we can obtain Versions~$B$ and $C$ as well as
the heat kernel $\sigma_{\mu, t}$ on $\mathbb{R}^N$ from Version $A$.
We also have the identities
\begin{eqnarray}
\label{A_mu_t}
A_{\mu, t} (z,q) &=& C_{\mu, t} (z,q) \, / \, ( C_{\mu, t} (0,q) )^{1/2}
\\
\label{fourth_mu_id}
  B_{\mu, t} (z,q) &=& C_{\mu, t} (z,q) \, / \, C_{\mu, t} (0,q)
\\
\label{fifth_mu_id}
\sigma_{\mu, t} (q) = \rho_{\mu, t} (0, q) &=& C_{\mu, t} (0,q)
\end{eqnarray}
which tell us that
we also can get Versions~$A$ and $B$ and the heat kernel $\sigma_{\mu, t}$
on~$\mathbb{R}^N$ from Version $C$.
\end{thm}
\begin{rem}
It is curious that in the present Coxeter context Version~$A$ determines Version~$C$
via the two distinct identities (\ref{third_mu_id}) and (\ref{Euclidean_id}).
We do not pretend to have any deeper understanding of this fact.
We will see that in the Lie group context only (\ref{third_mu_id}) has a valid analogue,
while the analogue of (\ref{Euclidean_id}) is false at least for the Lie group $SU(2)$.
\end{rem}
\textbf{Proof:}
For (\ref{first_mu_id}) we use (\ref{define_B}) and (\ref{rho_def}) to compute
\begin{equation}
\label{compute_B}
   B_{\mu, t} (z,q) = \dfrac{ \rho_{\mu, t}(z,q) }{ \rho_{\mu, t} (0,q) } = e^{-z^2/2t}
                      E_\mu \left(  \dfrac{z}{t^{1/2}}, \dfrac{q}{t^{1/2}} \right).
\end{equation}
We recall that $A_{\mu, t} (0,q) = e^{ - q^2/4t}$ which
together with (\ref{define_A}) implies that
\begin{equation}
\label{AoverA}
     \dfrac{A_{\mu, t} (z,q)}{A_{\mu, t} (0,q)} =  e^{-z^2/2t}
                      E_\mu \left(  \dfrac{z}{t^{1/2}}, \dfrac{q}{t^{1/2}} \right).
\end{equation}
Then equations (\ref{compute_B}) and (\ref{AoverA}) imply (\ref{first_mu_id}).

Next we note that (\ref{third_mu_id}) is exactly (\ref{C-A2}),
first proved in \cite{SBS2}.
To obtain (\ref{second_mu_id}) we put $z=0$ into (\ref{third_mu_id}).

We first proved (\ref{Euclidean_id}) in \cite{SBS2}.
This is a generalization of equation (A.18) in Hall's paper \cite{HA}, which
corresponds to the case $\mu \equiv 0$ of (\ref{Euclidean_id}).
This identity seems to be related to the fact that the underlying
Riemannian manifolds $\mathbb{R}^N$ and $\mathbb{C}^N$ are flat Euclidean spaces.

To prove (\ref{A_mu_t}) we calculate that
\begin{gather*}
 \dfrac{ C_{\mu, t} (z,q) }{ ( C_{\mu, t} (0,q) )^{1/2} }=
 \dfrac{ \rho_{\mu, t} (z,q) }{ ( \rho_{\mu, t} (0,q) )^{1/2} }=
 \dfrac{ \rho_{\mu, t} (z,q) }{ ( e^{-q^2/2t} )^{1/2} }=
  e^{q^2/4t}  \rho_{\mu, t} (z,q) =
\\
e^{q^2/4t}  e^{-(z^2 + q^2)/2t}
           E_\mu \left(  \dfrac{z}{t^{1/2}}, \dfrac{q}{t^{1/2}} \right) =
e^{-z^2/2t - q^2/4t} E_\mu \left(  \dfrac{z}{t^{1/2}}, \dfrac{q}{t^{1/2}} \right) =
A_{\mu, t} (z,q).
\end{gather*}

For (\ref{fourth_mu_id}) we merely note that by definition we have
$C_{\mu, t} (z,q) = \rho_{\mu, t} (z,q) $, and then we apply (\ref{compute_B}).
And finally we remark that (\ref{fifth_mu_id}) is just a special case of
$C_{\mu, t} (z,q) = \rho_{\mu, t} (z,q) $.
\hfill $\blacksquare$

\begin{rem}
Some of the results of Theorem \ref{Coxeter_ids},
such as (\ref{fifth_mu_id}), are well known,
while (\ref{third_mu_id}) is a relatively recent result.
We have presented all these identities together to emphasize the exact relations
among the three versions in the Coxeter case.
We will then use all this as motivation for the results in the next section.
\end{rem}

\section{Lie group case}

We now examine the corresponding case introduced by Hall
in \cite{HA} for a compact, connected (real) Lie group $K$.
We first review some material from \cite{HA} and refer the reader
to that paper for more details.
Now $K$ has a complexification, which is a complex Lie group $G$.
Among other things, $K$ is a Lie subgroup of $G$.
For every $t>0$ there is a heat kernel $ \rho_t : K \to (0,\infty) $, which
has a unique holomorphic extension (also denoted as $ \rho_t $)
with $\rho_t : G \to \mathbb{C}$.
We continue to consider $t>0$
in the following as Planck's constant and as having a fixed value.

The integral kernel function for Version $A$ is defined by
$$
     A_t (g,x) := \dfrac{\rho_t(x^{-1}g)}{(\rho_t(x))^{1/2}}
$$
for $g \in G$ and $x \in K$.
Here $x^{-1} g$ is in $G$ (but not necessarily in $K$)
and so the $\rho_t$ in $\rho_t(x^{-1} g)$
refers to the holomorphic extension.
The corresponding Version~$A$ Segal-Bargmann transform is then defined by
$$
          A_t \psi (g) := \int_{K} \mathrm{d}_{H} x \, A_t (g, x) \psi (x)
$$
for all $\psi \in L^2 (K, d_{H} x)$ and all $g \in G$, where $d_{H} x$
is the normalized Haar measure of the compact group $K$.
Theorem~1 in \cite{HA} states that $A_t : L^2 (K, d_{H} x) \to \mathcal{H} L^2 (G, \mu_t ) $
is a unitary isomorphism, where $\mu_t$ is a heat kernel measure on $G$
(and not to be confused
with our notation $\mu$ for the multiplicity function)
and $\mathcal{H} L^2 (G, \mu_t )$ denotes the closed subspace of holomorphic functions in
$L^2 (G, \mu_t ) $.

Theorem~2 in \cite{HA} states that $C_t : L^2 (K, d_{H} x) \to \mathcal{H} L^2 (G, \nu_t ) $
is a unitary isomorphism, where $\nu_t $ is the measure on $G$ that we get by averaging
$\mu_t$  over the left action of $K$ on $G$, using the fact that $K$ is a subgroup of $G$.
Of course, $\mathcal{H} L^2 (G, \nu_t )$ denotes
the closed subspace of holomorphic functions in $L^2 (G, \nu_t ) $.
The definition of the Version~$C$ Segal-Bargmann transform is
$$
          C_t \psi (g) := \int_{K} \mathrm{d}_{H} x \, C_t (g, x) \psi (x)
$$
for all $\psi \in L^2 (K, d_{H} x)$ and all $g \in G$, where the kernel function
is defined by
$$
     C_t (g,x) := \rho_t(x^{-1}g) 
$$
for $g \in G$ and $x \in K$.

However, for the $B$ Version we are using our convention (see \cite{SBS2}) that
$$
           B_t (g,x ) := \rho_t (x^{-1} g) \, / \, \rho_t (x)
$$
for $g \in G$ and $x \in K$,
which differs from the convention in \cite{HA}. 
In our convention a kernel function of two variables $T(x,y)$
determines an associated integral kernel transform $T$ by
$$
       Tf(x) := \int_Y \mathrm{d}\nu(y) \, T(x,y) f(y),
$$
where $(Y,\nu)$ is a measure space and $f$ is in a space
associated with the measure~$\nu$, say in $L^p (Y, \nu)$ for some $p$.
Note that we use the same symbol for the kernel function as well as for
its associated operator.
This is a common abuse of notation.

So our definition of Version $B$ Segal-Bargmann transform reads
$$
      B_t \phi (g) := \int_K \mathrm{ d } \rho_t(x) \, B_t (g,x) \phi (x),
$$
where $\mathrm{ d } \rho_t(x) := \rho_t(x) \, \mathrm{ d }_H x$,
for all $g \in G$ and $\phi \in L^2( K, \rho_t)$.
This is equivalent to the definition given in \cite{HA}.

The reader should note the analogy between this material from \cite{HA} and
our corresponding material in \cite{SBS2}, which was motivated by \cite{HA}.
In contrast, in this paper our results in the Coxeter context
given in the previous section will be used to motivate the study
of analogous results in the Lie group case.

Another analogy with the Coxeter case concerns the heat kernel.
In the Lie group context the heat kernel $\rho_t : K \to (0,\infty)$
determines two more kernels.
But first for each $x \in K$ and $f : K \to \mathbb{C} $ we define the
translation of $f$ by $x$ to be $ (\mathcal{T}_x f) (y) := f(x^{-1} y)$
for all $y \in K$.
This definition
has the virtue that $\mathcal{T}_{x_1} \mathcal{T}_{x_2} = \mathcal{T}_{x_1 x_2}$.
Then we define the two-variable heat kernel $\rho_t : K \times K \to (0,\infty)$
(using the same notation $\rho_{t}$ for this function) for $x,y \in K$ by
$$
    \rho_{t} (x,y) := (\mathcal{T}_x \rho_t)(y).
$$
This kernel in turn has an analytic continuation $\rho_{t} : G \times G \to \mathbb{C}$
(denoted again with the same notation), which is used in the definitions of
the kernel functions for all three versions of the Segal-Bargmann transform
in the Lie group context.

We would also like to note that there seems to be a limit as to how far one
can find analogies between the Coxeter case and the Lie group case.
For example, as noted above,
in the Lie group case the heat kernel measure of $G$ plays an important
role in defining the spaces of holomorphic functions on $G$.
However, even when $N=1$, the definition of the holomorphic function spaces
in the Coxeter case uses in general more than one measure.
(See \cite{SBS0}.)

We now are about ready to state our result for Lie groups.
But first we remark that $e$ denotes the identity element in $K\subset G$.
\begin{thm}
\label{Lie_thm}
Let $K$ be a compact, connected Lie group, and let $G$ denote its complexification.
Then we have for all for $g \in G$ and $x \in K$ the identities
\begin{eqnarray}
\label{first_id}
 B_t (g,x) &=& A_t(g,x) \, / \, A_t(e,x)
\\
\label{CAA}
   C_{t} (g,x) &=& A_{t} (e,x) A_{t}  (g,x)
\\
\label{second_id}
\rho_t(x) &=& ( A_t(e,x) )^2,
\end{eqnarray}
which tell us that from Version~$A$ we can obtain
Versions~$B$ and $C$ as well as the heat kernel $\rho_{t}$ on $K$
(and hence implicitly its analytic extension to $G$).
We also have the identities
\begin{eqnarray}
\label{zero_id}
A_t (g,x) &=& C_t(g,x) \, / \, ( C_t(e,x) )^{1/2}
\\
\label{fourth_id}
 B_t (g,x ) &=& C_t(g,x) \, / \,C_t(e,x)
\\
\label{fifth_id}
\rho_t (x) &=& C_t (e,x),
\end{eqnarray}
which tell us that
we can also get Versions~$A$ and $B$ and the heat kernel $\rho_{t}$ on $K$ from Version~$C$.
Finally, we have that
\begin{equation}
\label{conseq}
      \mathcal{H} L^2 (G, \mu_t ) = \mathcal{H} L^2 (G, \nu_t ).
\end{equation}
\end{thm}

\begin{rem}
All of the identities in Theorem \ref{Coxeter_ids} have an analogue here
except for equation (\ref{Euclidean_id}).
The identity (\ref{CAA}), which we believe to be new even though
it is quite elementary, shows that in this Lie group context the
Segal-Bargmann transform for Version~$A$ determines that for Version~$C$.
While it remains true that Version~$A$ and Version~$C$
are different (as in the Coxeter context),
there is an \textit{essential} relation between them and, indeed, a relation that
also holds analogously in the Coxeter context.
\end{rem}

\textbf{Proof:}
Let $g \in G$ and $x \in K$ be arbitrary in this proof.
For (\ref{CAA}) we simply use the definitions and $\rho_t(x^{-1}) = \rho_t(x)$
(see \cite{HA}, p.~108) to evaluate
$$
     A_{t} (e,x) A_{t}  (g,x) = \dfrac{\rho_t(x^{-1}e)}{(\rho_t(x))^{1/2}} \cdot
\dfrac{\rho_t(x^{-1}g)}{(\rho_t(x))^{1/2}} = \rho_t(x^{-1}g) = C_t (g,x).
$$
As in the Coxeter case, (\ref{CAA}) immediately implies a bounded inclusion, namely
$$
      \mathcal{H} L^2 (G, \nu_t ) \subset \mathcal{H} L^2 (G, \mu_t ),
$$
since $A_t(e,x) $ as a function of $x \in K$ is bounded, $K$ being compact.
By (\ref{CAA})
\begin{equation}
\label{ACA}
       A_{t}  (g,x) = \dfrac{C_{t} (g,x)}{ A_{t} (e,x) },
\end{equation}
which \textit{is} useful unlike (\ref{AC}).
This is so since the denominator satisfies
\begin{equation}
\label{Aex}
 A_{t} (e,x) = \dfrac{\rho_t(x^{-1}e)}{(\rho_t(x))^{1/2}} = (\rho_t(x))^{1/2} > 0,
\end{equation}
and so is bounded from below away from $0$, since $x$ varies in $K$ compact.
(For the inequality $\rho_t(x) > 0 $, see \cite{HA}.)
Given this fact, equation (\ref{ACA}) now implies that
$$
      \mathcal{H} L^2 (G, \mu_t ) \subset \mathcal{H} L^2 (G, \nu_t ),
$$
a bounded inclusion.
Together with the previous inclusion, this shows that
$$
      \mathcal{H} L^2 (G, \mu_t ) = \mathcal{H} L^2 (G, \nu_t ),
$$
which completes the proof of (\ref{conseq}).

The identity (\ref{second_id}) follows immediately from (\ref{Aex}).
Next, using the definitions of $A_t(g,x)$ and $B_t(g,x)$
as well as (\ref{Aex}), we calculate
$$
  \dfrac{A_t(g,x)}{A_t(e,x)} = \dfrac{\rho_t(x^{-1}g)}{(\rho_t(x))^{1/2}} \cdot
        \dfrac{1}{(\rho_t(x))^{1/2}} = \dfrac{\rho_t(x^{-1}g)}{\rho_t(x)} = B_t (g,x),
$$
which is exactly (\ref{first_id}).

To show (\ref{zero_id}) we simply note that
$$
  \dfrac{C_t(g,x)}{(C_t(e,x))^{1/2}} = \dfrac{\rho_t(x^{-1} g)}{(\rho_t(x^{-1}))^{1/2}}
 = \dfrac{\rho_t(x^{-1} g)}{(\rho_t(x))^{1/2}} = A_t(g,x).
$$

For (\ref{fourth_id}) we use $ C_t (e,x) = \rho_t (x^{-1}) = \rho_t (x) $
and definitions to get
$$
        \dfrac{C_t(g,x)}{C_t(e,x)} =  \dfrac{\rho_t(x^{-1}g)}{\rho_t(x)} = B_t (g,x),
$$
which proves (\ref{fourth_id}).
We have also just proved (\ref{fifth_id}), thereby finishing the proof.
\hfill $\blacksquare$

\begin{rem}
Of course, the heat kernel on $K$ determines all three versions of the
Segal-Bargmann transform, this being precisely a major theme of Hall's paper \cite{HA}.
The previous theorem shows that each of the Versions~$A$ and $C$ determines
the remaining two versions as well as the heat kernel of $K$.
\end{rem}

The identities (\ref{first_id})-(\ref{fifth_id}) are all easy to prove
and so it would be surprising if they are all new.
In fact, some of them clearly are not new, such as (\ref{fifth_id}).
However, the identity (\ref{CAA}) does seem to be new in this context.
But its consequence (\ref{conseq}) was already known,
since that follows from the stronger result
\begin{equation*}
      L^2 (G, \mu_t ) =  L^2 (G, \nu_t ),
\end{equation*}
which in turn follows immediately from Lemma~11 in \cite{HA} (p.~124).
However, (\ref{CAA}) gives us a quick, short proof of (\ref{conseq}).

The importance of (\ref{CAA}) is that it tells us Version~$C$ of the Segal-Bargmann
transform is determined by Version~$A$ of the Segal-Bargmann
transform, where we understand that an integral kernel transform is ``equivalent'' to
its integral kernel function.

We also wish to note that the inclusions and equalities of spaces
given in this section are \textit{as sets} and not as Hilbert spaces.
This is because the inner products do not coincide.

\section{An interesting counterexample: $SU(2)$}

As we have already remarked, the identity (\ref{Euclidean_id}) in the
Coxeter context did not have an analogue in the Lie group context.
We now construct a counterexample to show that the analogous equation
is false in general.
Much of the material in this section is classical.
We have chosen to start by following the presentation
and notation given in Chapter~7 of \cite{MI}.

We now consider the case of the compact Lie group $SU(2)$
of $2 \times 2$ complex matrices $A$
which are unitary (that is, $A^*A = I$) and have determinant one.
We use notation for group elements and the identity matrix (namely, $I$) that
is standard for matrix groups.
By the spectral theorem for normal operators, there exists a unitary matrix $B$ which
diagonalizes a given $A \in SU(2)$, that is $B^{-1} A B $ is diagonal.
By taking $C = B / ( \det B )^{1/2} \in SU(2)$, where $( \det B )^{1/2}$
is one of the square roots of $\det B$, we have that $A$ is conjugate in $SU(2)$ to
\begin{equation}
\label{first_diag}
           C^{-1} A C = \left(
\begin{array}{cc}
                          e^{\mathrm{i} \tau/2} & 0 \\
                             0  & e^{-\mathrm{i} \tau/2}
\end{array}
                         \right)
\end{equation}
for some real number $\tau \in [0, 4 \pi)$, since
the eigenvalues $\alpha, \beta$ of $A$ (and also of its diagonalization)
satisfy $ |\alpha| = |\beta| = 1$ and $ \alpha \beta = 1$.
The condition on $\tau$ is not too restrictive since it still allows
the $(1,1)$ entry (and also the $(2,2)$ entry)
in the matrix (\ref{first_diag}) to achieve any value on the unit circle.
But by conjugating formula (\ref{first_diag}) by the matrix
$$
\left(
\begin{array}{cc}
0 & \mathrm{i} \\
\mathrm{i} & 0
\end{array}
\right)
\in SU(2),
$$
which interchanges the eigenvalues on the diagonal of formula (\ref{first_diag}), we see
that the matrices in (\ref{first_diag}) with parameter $\tau \in [2 \pi , 4 \pi )$
are conjugate in $SU(2)$ to the matrices with parameter $\tau^{\prime} \in (0 , 2 \pi ]$,
where $\tau^{\prime} = 4 \pi - \tau $.

Moreover, for $\tau_{1}, \tau_{2} \in [0 , 2 \pi ]$ with $\tau_{1} \ne \tau_{2}$, the
corresponding matrices are not conjugate, since they have different sets of eigenvalues.
So, by taking $\tau \in [0, 2 \pi]$ in (\ref{first_diag}), we get a family of matrices
which contains exactly one representative of each conjugacy class in $SU(2)$, that is,
the value of $\tau$ in $[0, 2\pi]$ is now uniquely determined for each $A \in SU(2)$.

Even though we could label the irreducible unitary representations of $SU(2)$ by
their dimensions, it is conventional to label them by the non-negative half
integers $u$  (those non-negative real numbers $u$ such that $2u$ is an integer)
such that $2u+1$ is the dimension of the representation.
If $\phi_u$ denotes the associated irreducible representation,
then we have that $\phi_u(A)$ is a $(2u+1) \times (2u+1)$
unitary matrix for every $A \in SU(2)$.
The corresponding character $\chi_u = Tr \circ \phi_u$ (where $Tr$
is the trace of a matrix) is a complex-valued function
that is constant on each conjugacy class of $SU(2)$.
So, $\chi_u (A)$ is a function of $\tau \in [0, 2\pi]$ only.
Actually, this function can be calculated explicitly for $A \in SU(2)$ as
$$
\chi_u (A) = \sum_{s=-u}^u e^{is\tau} = \dfrac{\sin(u+1/2)\tau}{\sin (\tau/2)},
$$
where these formulas can be found in \cite{MI}, p.~232.
The last formula results by summing the finite geometric series and simplifying.
(The singularities in the last expression are removable
and are understood as having been removed.)
Note that the summation in the second expression is taken in unit steps,
even in the case when $u$ is not an integer.
For example, when $u=3/2$ the sum is over $s$ equal to the four values
$-3/2$, $-1/2$, $1/2$, $3/2$.
In general, the sum contains $2u+1$ terms.

We have used \cite{MI} as a guide for the discussion so far but now take a different tack,
since we wish to write $\tau$ in terms of the matrix $A$.
Note that any $A \in SU(2)$ can be written as
$$
     A= \left(
               \begin{array}{cc}
                   a & b \\
                   -b^* & a^*
                \end{array}
        \right)
$$
with $a,b \in \mathbb{C}$ satisfying $|a|^2 + |b|^2 =1$.
Therefore we have that
\begin{equation}
\label{TrA}
Tr(A) = a + a^* = 2 Re (a)
\end{equation}
and by (\ref{first_diag}) that
$$
Tr(A) = Tr (C^{-1} A C) = e^{\mathrm{i} \tau/2} + e^{-\mathrm{i} \tau/2} = 2 \cos(\tau/2).
$$
So we have $Re (a) = \cos (\tau/2)$ or equivalently
$$
        \tau = 2 \cos^{-1} ( Re(a) ).
$$
Here we are using the standard definition $\cos^{-1} : [-1, 1] \to [0,\pi]$.
We note that $Re(a) \in [-1,1]$, since $|Re(a)|^2 \le  |a|^2 \le |a|^2 + |b|^2 =1$.
So this is in agreement with our earlier restriction that $\tau \in [0, 2\pi]$,
that is, our choice for the branch of the inverse cosine is correct.

Returning to the character, we see that
$$
\chi_u (A) = \dfrac{\sin(u+1/2)\tau}{\sin (\tau/2)} =
              \dfrac{\sin [(2u+1) \cos^{-1} ( Re(a) ) ] }{\sin [ \cos^{-1} ( Re(a) ) ] },
$$
which already expresses the character of $A \in SU(2)$ in terms of an entry of
the matrix $A$, namely $a$, although the formula seems to leave something to be desired.

Now we recall the definition of the \textit{Chebyshev polynomial of the second kind}
(see \cite{NIK})
for
any $x \in (-1,1)$ and integer $n \ge 0$ as
$$
  U_n(x) := \dfrac{\sin((n+1)\theta)} {\sin \theta} =
            \dfrac{\sin((n+1)\cos^{-1}x) } {\sin ( \cos^{-1} x ) },
$$
where $x=\cos \theta$ or $\theta = \cos^{-1} x$.
One verifies that this is a polynomial function of~$x$ in the interval
$(-1,1)$ and then extends the domain of definition of $U_n$ to the entire
complex plane by analytic continuation.

So, for $|Re (a)| < 1$, we finally arrive at the rather simple expressions
\begin{equation}
\label{simple_express}
   \chi_u (A) = U_{2u} ( Re(a) ) = U_{2u} \big( \dfrac{1}{2} Tr (A) \big),
\end{equation}
where the second equality comes from equation (\ref{TrA}).
(The case $|Re (a)| = 1$ occurs if and only if $A = \pm I $.
Then the proof of (\ref{simple_express}) follows by continuity.)
Now these are more elegant ways of writing the character of $A \in SU(2)$ in terms of
$A$ itself.
We must note here that the formula (\ref{simple_express}) is known, but
apparently not that well appreciated.
For example, Miller notes in \cite{MI}, p.~233, that he is aware of this formula,
but he does not present it since he considers it to be ``not very enlightening.''
This is why we have presented and proved (\ref{simple_express}) here.

Next, according to equation (15) in \cite{HA}, the heat kernel
of the compact Lie group $SU(2)$ is given for $A \in SU(2)$ and $t>0$ by
$$
\rho_t (A) = \sum_u \dim (\phi_u) e^{-\lambda_{u}t/2} \, \chi_u (A),
$$
where $\dim (\phi_u) = 2u + 1$ and $\lambda_{u}$ is the \textit{unique} eigenvalue
of minus the Laplacian acting in the representation space, that is
$\phi_u (- \Delta ) = \lambda_u I$.
We have $\lambda_u = u(u+1)$.
(See \cite{MI}.)
So we have  that
\begin{gather}
\rho_t (A) = \sum_u (2u+1) e^{-u(u+1)t/2} \, U_{2u} \big( \dfrac{1}{2} Tr (A) \big)
\nonumber
\\
\label{rho_t_A}
= \sum_{n=0}^\infty (n+1) e^{-n(n+2)t/8} \, U_n \big( \dfrac{1}{2} Tr (A) \big),
\end{gather}
where the first sum is over all non-negative half-integers $u$ and the second is over all
integers $n \ge 0$, where $n = 2u$.
Hall proves in \cite{HA} that this series converges absolutely
for all $t>0$ and all $A \in SL(2; \mathbb{C})$.
However, by using properties of the polynomials $U_n$, one can directly prove the
absolute convergence of this series for all $t>0$
and for \textit{any} $2 \times 2$ complex matrix $A$, as we will show momentarily.
Also, it is known that $\rho_t (A) >0$ for all $t >0$ and 
for all $A \in SU(2)$, but this is not obvious from formula (\ref{rho_t_A}), since
$U_n$ has $n$ simple roots in $[-1, 1]$.

   We now consider how to find the analytic continuation of the heat kernel $\rho_t$ to the
complexification of $SU(2)$, which also can be identified with the cotangent
bundle of $SU(2)$.
It turns out that the complexification of $SU(2)$ is $SL(2; \mathbb{C})$,
the group of all $2 \times 2$ matrices with complex entries and determinant one.
We will next show that the analytic continuation of $\rho_t$
is given by the same formula (\ref{rho_t_A})
given above, but now for $A \in SL(2; \mathbb{C})$.
Actually, we will procede by proving the uniform absolute convergence of the series in
 (\ref{rho_t_A})
on compact subsets of $M(2; \mathbb{C})$, the space of all
$2 \times 2$ complex matrices.
First, we need a lemma.

\begin{lem}
For all $z \in \mathbb{C}$ and every integer $k \ge 0 $ we have
this estimate for the Chebyshev polynomials $U_k$ of the second kind:
$$
 | U_k (z) | \le ( \, 3 \max ( 1, |z| ) \, )^k
$$
\end{lem}
\begin{rem}
This estimate is not optimal.
Nor is it meant to be.
\end{rem}
\textbf{Proof:}
The proof is by induction on $k$.
For $k=0$ and $k=1$ the estimate is easy enough, using $U_0(z) =1$ and $U_1(z) = 2z$,
and so is left to the reader.
We now assume that $n \ge 1$ and that the estimate holds for
$k = n $ and $ k = n-1$.
It remains for us to show the estimate for $k = n+1$.
We will use the three term recursion relation
for the Chebyshev polynomials of the second kind:
$$
        U_{n+1} (z) = 2 z U_n (z) - U_{n-1} (z)
$$
for $n \ge 1$. (See \cite{DX}.)
We first consider the case $|z| \ge 1$.
Using the induction hypothesis we have that
\begin{eqnarray*}
 | U_{n+1} (z) | &\le& 2 |z| \,  |U_n (z)| + | U_{n-1} (z) |
\le 2 |z| \,  3^n |z|^n + 3^{n-1} |z|^{n-1}
\\
&\le& 2 \cdot 3^n |z|^{n+1} + 3^{n-1} |z|^{n+1} = (2 \cdot 3 +1 ) \, 3^{n-1} |z|^{n+1}
\\
&\le& 3^2 \, 3^{n-1} |z|^{n+1} = 3^{n+1} |z|^{n+1},
\end{eqnarray*}
which is the estimate for $n+1$ in this case.
The case $|z| \le 1$ is proved similarly.
\hfill $\blacksquare$

\begin{thm}
\label{cau}
The series in (\ref{rho_t_A}) converges absolutely and
uniformly on compact subsets of $M(2; \mathbb{C})$.
Consequently, it defines a holomorphic function on the complex manifold
$M(2; \mathbb{C}) \cong \mathbb{C}^{4}$.
\end{thm}
\textbf{Proof:}
Consider a compact subset $S \subset M(2; \mathbb{C}) $.
Define
$$
C_S := 3 \max( \, 1, \sup_{A \in S} \dfrac{1}{2} |Tr (A) | \, ).
$$
Then by the previous lemma we have that $ | U_n ( \frac{1}{2} Tr(A) ) | \le  (C_S)^n$ for
all $A \in S$.
We use this and the root test to estimate for $A \in S$ as follows:
$$
 \sum_{n=0}^\infty \big| (n+1) e^{-n(n+2)t/8} \, U_n \big( \dfrac{1}{2} Tr (A) \big) \big|
\le
 \sum_{n=0}^\infty  (n+1) e^{-n(n+2)t/8} \,  (C_S)^n < \infty.
$$
The first statement of the theorem now follows from the Weierstrass $M$-test.
Since $Tr : M(2; \mathbb{C}) \to \mathbb{C}$ is clearly holomorphic and
each $U_n$ is a holomorphic polynomial, the
partial sums of (\ref{rho_t_A}) are clearly holomorphic functions
of $A \in M(2; \mathbb{C})$.
So the second statement of the theorem follows immediately from the first statement.
\hfill $\blacksquare$

\begin{rem}
Since the inclusion mapping $SL(2; \mathbb{C}) \hookrightarrow M(2; \mathbb{C}) $
is holomorphic, it follows that (\ref{rho_t_A})  for $A \in SL(2; \mathbb{C})$
gives the analytic continuation of $\rho_t$ from $SU(2)$ to $SL(2; \mathbb{C})$.
Theorem \ref{cau} follows from \cite{HA} (Prop.~1, p.~111)
but only for $SL(2; \mathbb{C})$ instead of $M(2; \mathbb{C})$.
\end{rem}

We now change notation by letting $X \in SU(2)$ and $G \in SL(2; \mathbb{C})$
denote generic elements in these two groups.
Then the integral kernel for Version~$A$ of the Segal-Bargmann transform
for $SU(2)$ is given by
$$
A_t (G,X) = \dfrac{\rho_t(X^{-1}G)}{(\rho_t(X))^{1/2}},
$$
while the kernel for Version~$C$ of the Segal-Bargmann transform for $SU(2)$ is
$$
C_t (G, X) = \rho_t(X^{-1}G).
$$

Finally, we now prove the main result of this section, namely
that the identity analogous to (\ref{Euclidean_id}) is not true for $SU(2)$.
\begin{thm}
The equation
\begin{equation}
\label{can_not_be_true}
 C_{t}(G^2,X) = A_{2t} (G^2,I) A_{t/2}  (G,X)
\end{equation}
is not identically true for all $X \in SU(2)$ and all $G \in SL(2; \mathbb{C})$.
\end{thm}
\textbf{Proof:}
Equation (\ref{can_not_be_true}) is equivalent to
\begin{equation}
\label{counterexample}
\rho_{t} (X^{-1} G^2) = \dfrac{ \rho_{2t}(G^2) }{ ( \rho_{2t} (I) )^{1/2} } \cdot
\dfrac{ \rho_{t/2}(X^{-1}G ) }{ ( \rho_{t/2} (X) )^{1/2} }.
\end{equation}
Let us suppose that this is an identity and try to get a contradiction.
First we calculate the heat kernel of two elements of $SU(2)$.
For the identity matrix $I$ we have that
\begin{eqnarray*}
\rho_t (I)
&=&
 \sum_{n=0}^\infty (n+1) e^{-n(n+2)t/8} \, U_n \big( \dfrac{1}{2} Tr (I) \big)
\\
&=&
\sum_{n=0}^\infty (n+1) e^{-n(n+2)t/8} \, U_n ( 1 )
=\sum_{n=0}^\infty (n+1)^2 e^{-n(n+2)t/8},
\end{eqnarray*}
where we used that $U_n(1) = n + 1$ and $Tr(I)=2 $.
For the next calculation we use $-I \in SU(2) $ and $U_n(-1) = (-1)^n (n + 1)$.
We then have
\begin{eqnarray*}
\rho_t (-I)
&=&
 \sum_{n=0}^\infty (n+1) e^{-n(n+2)t/8} \, U_n \big( \dfrac{1}{2} Tr (-I) \big)
\\
&=&
\sum_{n=0}^\infty (n+1) e^{-n(n+2)t/8} \, U_n ( -1 )
=\sum_{n=0}^\infty (-1)^n (n+1)^2 e^{-n(n+2)t/8}.
\end{eqnarray*}
This clearly implies that $\rho_t (-I) < \rho_t (I)$.
We also have $ 0 < \rho_t (-I)$ by the strict positivity of
the heat kernel on $SU(2)$.

Next in (\ref{counterexample}) we take $X=I$ and $G=I$ to get
$$
\rho_{t} (I) = \dfrac{ \rho_{2t}(I) }{ ( \rho_{2t} (I) )^{1/2} } \cdot
\dfrac{ \rho_{t/2}(I ) }{ ( \rho_{t/2} (I) )^{1/2} }.
$$

We also take $X=-I$ and $G=-I$ in (\ref{counterexample}) thereby obtaining
$$
\rho_{t} (-I) = \dfrac{ \rho_{2t}(I) }{ ( \rho_{2t} (I) )^{1/2} } \cdot
\dfrac{ \rho_{t/2}(I ) }{ ( \rho_{t/2} (-I) )^{1/2} }.
$$
Note that in these last two equations all the values of the heat kernel
are strictly positive real numbers.
So, it follows that
$$
0 \, < \, \rho_{t} (-I) (\rho_{t/2} (-I) )^{1/2} \, = \,\rho_{t} (I) (\rho_{t/2} (I) )^{1/2}.
$$
And this contradicts $0 < \rho_t (-I) < \rho_t (I)$, which holds for \textit{all}
$t>0$.
\hfill $\blacksquare$

\begin{rem}
There are surely many other ways to show that (\ref{counterexample}) is not an identity.
For example, one could use a computer assisted proof.
\end{rem}

\section{Concluding remarks}

We feel that a major result of this note is embodied in (\ref{CAA}), which tells us that
Version~$C$ is determined by Version~$A$ in the Lie group context.
Moreover, (\ref{zero_id}) gives us the reciprocal relation that
Version~$A$ is determined by Version~$C$ in the Lie group context.
It was our study of Segal-Bargmann analysis in the Coxeter context which motivated us
to find these results.

It is reasonable to conjecture that (\ref{can_not_be_true}) is false for
every compact, connected, non-abelian Lie group $K$.
In the contrary case it would be interesting to know for which such $K$
equation (\ref{can_not_be_true}) is an identity.

The multitude of analogies between the Segal-Bargmann theory associated to a
Coxeter group and the Segal-Bargmann theory for compact Lie groups strongly suggests
that there is more here than mere analogy.
An avenue for further research would be to find out if there is for example
a new general theory which has these two theories as special cases.
However, there is a difference, which we would like to note, between these two theories.
The Lie group case is based on a Laplacian associated to the Lie group.
And this is a differential operator.
But the Coxeter case is based on the Dunkl Laplacian,
which is a differential-difference operator.

\subsection*{Acknowledgment}
I wish to thank Brian Hall for various comments that clarified some matters 
in this paper for me.

\end{document}